# A semantic approach to mapping the Provenance Ontology to Basic Formal Ontology

## Authors

Tim Prudhomme[a,b], Giacomo De Colle[a,c], Austin Liebers[a,c], Alec Sculley[a,d], Peihong "Karl" Xie[a,e], Sydney Cohen[a] and John Beverley[a,c,f]

[a] National Center for Ontological Research, Buffalo, NY, USA

[b] Conceptual Systems, Chapel Hill, NC, USA

[c] Department of Philosophy, University at Buffalo, Buffalo, NY, USA

[d] Summit Knowledge Solutions, Arlington, VA, USA

[e] Department of Philosophy, University of Vienna, Vienna, Austria

[f] Institute for Artificial Intelligence and Data Science, Buffalo, NY, USA

## Abstract

The Provenance Ontology (PROV-O) is a World Wide Web Consortium (W3C) recommended ontology used to structure data about provenance across a wide variety of domains. Basic Formal Ontology (BFO) is a top-level ontology ISO/IEC standard used to structure a wide variety of ontologies, such as the OBO Foundry ontologies and the Common Core Ontologies (CCO). To enhance interoperability between these two ontologies, their extensions, and data organized by them, a mapping methodology and set of alignments are presented according to specific criteria which prioritize semantic and logical principles. The ontology alignments are evaluated by checking their logical consistency with canonical examples of PROV-O instances and querying terms that do not satisfy the alignment criteria as formalized in SPARQL. A variety of semantic web technologies are used in support of FAIR (Findable, Accessible, Interoperable, Reusable) principles.

## Introduction

The *data silo problem* is the problem of having data stored in isolated databases all of which are organized according to independent schemes [1]. Data sets in different data silos are not interoperable. Ontologies are well-structured vocabularies that logically define classes and relationships in the interest of promoting interoperability [2]. A popular way to construct ontologies, and the way relevant to this paper, is by leveraging the W3C standards Web Ontology Language (OWL) [3] and Resource Description Framework (RDF) [4]. Data sets become more semantically interoperable when interpreted by the same OWL ontology into a *knowledge graph* [5]. While ontologies are a promising strategy for remedying the data silo problem, the proliferation of ontologies in a domain may lead to larger ontology silos. The *ontology silo problem* is the problem of having data organized by independent ontologies. Although a data set organized within an ontology is internally interoperable, different data sets organized by independent ontologies are not mutually interoperable. One solution is to create mappings between terms in ontologies of interest.

An *ontology mapping*, or correspondence, is a statement <s, p, o> such that 's' is a subject term representing a class or object property in a ontology, 'o' is an object term representing a class or object property in some other ontology, and 'p' is a predicate that specifies how s and o relate [6, 7]. An *ontology alignment* is a set of ontology mappings [6]. Data sets tagged with aligned ontologies are more or less semantically interoperable based on the number of mappings and types of predicates used in the mappings. Maximum semantic interoperability can be achieved via a *synonymous alignment*, where every term in both ontologies is mapped using only predicates representing equivalence relations. The methodology we present outlines theoretical criteria and practical techniques for maximizing the interoperability of ontology alignments in stages which culminate in synonymy. Our alignments are encoded in machine-readable axioms that provide a basis for automated reasoning over data tagged by two ontologies. In this way, human intervention is minimized, and two ontologies can be automatically used together consistently.

The Provenance Ontology (PROV-O) is a W3C recommended ontology used to structure data about provenance: "information about entities, activities, and people involved in producing a piece of data or thing, which can be used to form assessments about its quality, reliability or trustworthiness" [8, 9]. Basic Formal Ontology (BFO) is a top-level ontology ISO standard [10] used to provide foundational classes to structure different domain ontologies and to allow for semantic interoperability between them [11, 12]. The OBO Relations Ontology (RO) is an extension of BFO developed by Open Biomedical Ontologies (OBO) Foundry [13, 14, 15] to standardize relations between domain ontologies. The Common Core Ontologies (CCO) are a suite of mid-level ontologies, used to span across different domain ontologies and intended to bridge the gap between domain ontologies and BFO [16, 17].

Given the success of BFO as a top-level ontology to integrate data tagged with different ontologies and the widespread adoption of PROV-O as an ontology for provenance, a successful alignment between the two would provide integration between all the knowledge graphs currently structured according to BFO and PROV-O, as well as homogeneity between the recommended standards for semantic interoperability.

In what follows, we apply our methodology by mapping PROV-O and its extensions PROV-AQ, PROV-Dictionary, PROV-Links, PROV-Inverses, and PROV Dublin Core to BFO and two of its extensions: CCO and RO. The result is a total alignment of every class and object property in all W3C extensions of PROV-O to some class or object property in BFO, CCO, or RO, using subsumption and equivalence relation mappings. The data files which encode the alignments are maintained at https://github.com/BFO-Mappings/PROV-to-BFO and archived at https://doi.org/10.5281/zenodo.14692262 [18].

Our alignments are designed such that users of BFO-conformant ontologies will be able to structure data about provenance using PROV-O terms. This allows for terms from PROV-O to be used in BFO conformant domain ontologies in order to facilitate interoperability in information systems and to support querying and reasoning over data. Our project's ultimate aim is to allow for new inferences to be drawn from the integration of extra semantic information coming from other ontologies. For example, new SPARQL queries could be run over PROV-O data also using BFO classes, and a semantic reasoner using axioms of BFO could discover new implicit information in PROV-O conformant knowledge graphs. Most significantly, our alignments were used to automatically detect minor mistakes in canonical PROV-O example data that are consistent with PROV-O alone.

## Methods

Our approach to mapping is based on the semi-automated curation of ontologies leveraging conceptual analysis techniques and semantic web technologies. This involves carefully evaluating the necessary and sufficient conditions for something to qualify as an instance of a class or relation represented in an ontology. Although this is a challenging process compared to fully-automated ontology matching techniques, we believe it is well-worth the effort given the generality and wide-spread use of both PROV-O and BFO. Our methodology can be described first by the theoretical criteria we chose for a successful alignment, and second by the engineering techniques used for rigorously evaluating an alignment and ensuring that its technical artifacts conform to FAIR principles [19].

To enhance legibility, the following writing conventions are used. Bold letters introduce terms in the context of a mapping. Italic letters are used for RDF properties, OWL object properties and for emphasis. We use the exact RDFS *label* of a term, including its capitalization, to either use or mention it. To distinguish between different terms with the same lexical label, we prefix the namespace of its ontology unless the ontology is implied by context. We refer to terms from PROV-O and its extensions with the namespace "PROV". For example: the PROV-O term PROV Activity represents the class of activities. We use the word "alignment" for a full set of mappings between two ontologies and use the word "mappings" for nonspecific subsets of an alignment.

### Criterion 1: Types of Mapping Relations

*Equivalence relations* represented by **OWL *equivalentClass*** and **OWL *equivalentProperty*** give necessary and sufficient conditions for something to be an instance of a certain BFO class or relation and a certain PROV-O class or relation at the same time. Everything that satisfies these conditions will be an instance of both, and nothing else will be an instance of either. An equivalence mapping provides a two-way bridge which allows for interoperability between two knowledge graphs. For example, the class PROV Activity is mapped as equivalent to the class BFO process. All instances of BFO processes are instances of PROV activities and vice versa, and both classes would stand in the same relation to other classes represented in a knowledge graph that used both ontologies together. If that is the case, then respective instances will now be differently axiomatized and produce new results when a reasoner is run over the knowledge graph that contains them.

*Subsumption relations* represented by **RDFS *subClassOf*** or **RDFS *subPropertyOf*** give sufficient conditions for an instance of one class or relation to be an instance of another class or relation [20]. If a certain PROV-O class is a subclass of a certain BFO class, then all the instances of the PROV-O class are also instances of the corresponding BFO class. The result of a subsumption relation mapping is then a one-way bridge from one term to another. Notice that the use of subsumption relations allows for *non-injective* alignments, such that multiple terms from one ontology can be subsumed under a single term from the other ontology.

Complex equivalence and subsumption relations between combinations of terms may be represented with **OWL *unionOf***, **OWL *intersectionOf***, and OWL property restrictions. SWRL rules [21] are especially useful for restricting the domain or range of an OWL object property in order to use it in a valid mapping. An advantage of SWRL is that it is implemented by semantic reasoners such as HermiT [22, 23]. SPARQL [24] queries can also encode mappings that are equivalent to some SWRL rules. If a set of relations in one ontology should imply a relation in another ontology, OWL property chain axioms can be used to axiomatize this complex subsumption relation between object properties.

Mapping predicates from the SKOS vocabulary [25] represent informal relations between terms which may be interpreted by users to be intuitively similar. SKOS is commonly used for ontology mappings, especially in conjunction with annotations from the Simple Standard for Sharing Ontology Mappings (SSSOM) vocabulary [7]. However, SKOS predicates have weaker inferential semantics than those of subsumption and equivalence in RDFS and OWL. For example, **SKOS *relatedMatch*** is a symmetric, non-transitive object property "used to state an associative mapping link between two conceptual resources in different concept schemes". A semantic reasoner will not infer anything about instances of two classes or object properties mapped using SKOS *relatedMatch* other than their symmetric relationship. This can be desirable when SKOS is used to both compare and contrast terms by providing a text description annotation of the mapping.

### Criterion 2: Coherence and Consistency

An ontology may be implemented as a logical theory, or a consistent set of formulae in a first-order formal language. The OWL implementations of BFO and PROV-O are based on first-order logic fragments which balance expressivity and computational efficiency for automated reasoning tasks such as classification or checking satisfiability. BFO, RO, and CCO conform to the OWL 2 DL profile, corresponding to the description logic SROIQ [3]. PROV-O and its extensions conform to OWL 2 RL, corresponding to the description logic DLP, with the exception of some axioms that conform to OWL 2 DL [8]. Our remaining criteria for evaluating ontology alignments and semantic interoperability are based on model-theoretic semantics [6, 26], which define the meaning of formal languages in terms of their possible interpretations. When a formula $\phi$ is true in every interpretation where formulae $\Gamma$ are true, we say that "$\Gamma$ entails $\phi$" or "$\Gamma \vDash \phi$".

An ontology alignment is *coherent* if and only if all formulae in the aligned ontologies are satisfiable, i.e. it is possible for every formula to be true in some interpretation (a model) [6]. An incoherent alignment might contain classes that cannot have instances. For example, a formula mapping PROV Plan as a subclass of both BFO continuant and occurrent is unsatisfiable: nothing could be an instance of a PROV Plan in that case because nothing can be both a continuant and occurrent in BFO. Such an instance would contradict this disjointness axiom in BFO, making the alignment also inconsistent. By contrast, a *consistent* alignment is free of entailed or derivable contradictions. An alignment should be both coherent and consistent, although checking for consistency requires instances.

**Coherence:** Let OM be the union of ontology O1, ontology O2, and an alignment between O1 and O2. The alignment is *coherent* if and only if every formula in OM is satisfiable.

**Consistency**: An ontology alignment is *consistent* if and only if the aligned ontologies entail no contradictions.

### Criterion 3: Conservativity

The *conservativity principle* [27] states that an ontology alignment should not change the semantic relationships between terms *within* each ontology. This is based on the notion of a conservative extension of a logical theory, which does not introduce or eliminate entailments expressed in the signature of that theory [28]. The signature, or vocabulary, of an OWL ontology is the set of terms representing its classes, object properties, data properties, individuals, datatypes, and literals [29]. It is undecidable whether an OWL 2 DL ontology is a conservative extension in the general sense [28]. A weaker, more tractable version of the conservativity principle aims to prevent changes to the subsumption relationships entailed by formulae of an ontology's signature. Solimando *et al.* [30] call the set of subsumptions, for a given ontology signature, entailed in one ontology but not in another, the *approximate deductive difference* between the ontologies. We use this to define a conservative alignment.

**Conservativity:** Let OM be the union of ontology O1, ontology O2, and an alignment between O1 and O2. The ontology alignment is *conservative* if and only if the approximate deductive difference between O1 and OM is empty for the signature of O1, and the approximate deductive difference between O2 and OM is empty for the signature of O2.

Informally, a conservative alignment in this sense does not change the meaning of terms *within* either ontology, at least with respect to how those terms are related in a subsumption hierarchy. As a counterexample, mapping PROV Entity as equivalent to BFO continuant, while mapping PROV Agent as a subclass of BFO continuant, would entail that PROV Agent is a subclass of PROV Entity. It may be true that some agents are entities in PROV-O, but it is not an axiom that all are. This new, entailed subsumption would be included in the approximate deductive difference between PROV-O and the union of PROV-O and its alignment to BFO, and thus violate conservativity in this sense.

### Criterion 4: Scope of Alignment

An ontology alignment can be characterized by the number of mappings between ontologies and the logical implications of those mappings. We describe three related types of alignments to provide context for our project. An ontology alignment can be defined for subsets of terms or formulae, which Euzenat [6] calls the “matchable entities”. For simplicity, we restrict our first alignment criterion to terms. When every term in one ontology is surjectively mapped onto by at least one term in the other ontology, this constitutes a *total alignment* [6].

**Totality:** An alignment $\Delta$ is a *total alignment* of an ontology O1 to ontology O2 for their signatures $\Sigma(O1)$ and $\Sigma(O2)$ if and only if for any term $\sigma \in \Sigma(O1)$, there exists some term $\varphi \in \Sigma(O2)$ and some relation $r \in \Theta$ where $\Theta$ is a set of mapping relation types, such that $\langle \sigma, r, \varphi \rangle \in \Delta$.

A total alignment is a syntactic relationship that does not necessarily translate the meaning of terms in an ontology. For an alignment to represent a semantic translation of an ontology, it should preserve the logical implications of the ontology after mapping each term, beyond conservatively extending it. More specifically, when every type of mapping relation in a total alignment is an equivalence relation, and the alignment is conservative, this is similar to the notion of *interpreting* one logical theory into another. An equivalence relation mapping which conservatively extends a logical theory is called a *translation definition* [31]. Informally, a translation definition provides a way of understanding a part of one theory in terms of another theory. Following Gruninger *et al.* [31], we can apply the same idea to ontologies.

**Interpretability:** An ontology O1 is *interpretable in* ontology O2 if and only if there exists a set of translation definitions (an alignment) $\Delta$ for O1 into O2 such that $O2 \cup \Delta \models O1$. In other words, any logical implication of the interpreted ontology also follows from the other ontology when combined with the alignment.

Informally, this means that one ontology can be translated into the other with no loss in meaning. If PROV-O were interpretable in BFO, then any data represented with PROV-O could be equivalently represented with BFO with at least the same logical implications of PROV-O. For example, PROV Activity is disjoint with PROV Entity. Our alignment maps these classes to BFO process and BFO continuant, respectively, which are also disjoint. The alignment combined with BFO entails the disjointness of the PROV-O classes, preserving the corresponding axiom in PROV-O.

As another example, the creation of a health record could be represented with PROV *wasGeneratedBy*. The meaning of this term implies that the record creation is a PROV Activity. Next we could check if our alignment preserves this implication. If we replace PROV *wasGeneratedBy* with BFO *participates in*, this analogously implies that the record creation is a BFO process. Since BFO process is mapped as equivalent to PROV Activity, this implies the record creation is a PROV Activity, preserving the original implication. However, because PROV *wasGeneratedBy* is mapped as a subproperty of BFO *participates in*, this does not imply that the relation is equivalent to a PROV *wasGeneratedBy* relation. So, interpretability would fall short here.

The most semantically interoperable type of alignment makes interpretability between ontologies bidirectional, demonstrating a kind of synonymy between their theories [32].

**Synonymy:** Two ontologies O1 and O2 are *synonymous* if and only if there exist two sets of translation definitions $\Delta$ and $\Pi$, respectively from O1 to O2 and from O2 to O1, such that $O1 \cup \Pi$ is logically equivalent to $O2 \cup \Delta$. In other words, each ontology entails the other when combined with their respective alignments.

Informally, this means that everything in either ontology can be understood as equivalent to something in the other ontology. A synonymous alignment provides the most interoperability between ontologies because each can be fully understood in terms of the other. It represents the theoretical limit of semantic interoperability. As Schorlemmer & Kalfoglou [33] write: "Two systems are semantically interoperable if [...] all implications made by one system hold and are provable by the other, and that there is a logical equivalence between those implications". Because synonymy involves a kind of bidirectional interpretation, we summarize our overall approach with the slogan "interpretability enhances interoperability".

### Summary: Criteria for Mapping PROV-O to BFO

Although PROV-O is a generic ontology, BFO represents domains that are not represented in PROV-O, such as those covering spatial and temporal regions. Thus, not every term in BFO can be mapped to some term in PROV-O. A total alignment of BFO to PROV-O, much less a synonymous alignment, is not possible. However, as an upper ontology, the domain of BFO includes the domain of PROV-O, so a total alignment of PROV-O to BFO with consistent logical implications is possible. But PROV-O also includes domain-specific terms related to provenance that may not be translatable to BFO or its extensions. In such cases where the domains overlap but the PROV-O term is more specific, a subsumption relation mapping between the BFO and PROV-O term can be used.

Given these considerations, we aimed to create at least a **coherent, consistent, conservative, total alignment of all classes and object properties in all extensions of PROV-O to some class or object property in an extension of BFO using only equivalence or subsumption mappings**, which could also be implemented with property chain axioms or SWRL rules. Data properties were not prioritized, due to representational limits of OWL and RDF, but we describe informal, complex mappings of these. This type of alignment maximizes interoperability with mappings that produce new inferences when the ontologies are used together. It also brings us closer to a full interpretation of PROV-O into extensions of BFO, as described in our **Interpretability** criterion. Additional SKOS mappings could add extra metadata about alternative mappings and informal relationships between terms, but we did *not* include SKOS relations as part of the formal alignment criteria used for evaluation, described in the next section.

Finally, explicit mappings for terms are *not* always included in cases where the relations for these mappings are *entailed* by logical axioms. For example, terms that are subsumed by a more generic term for which there is a mapping are not explicitly mapped due to the transitivity of subsumption, unless there are more specific terms that could appropriately map to those subclasses. For example, PROV Agent is mapped as a subclass of BFO material entities that participate in some PROV Activity. This implies that every instance of any PROV Agent subclass is an instance of that subclass of BFO material entity. Subclasses of PROV Agent therefore need not be explicitly mapped. Mappings for specific terms that inherit from more generic mapped terms are only included for more interoperability when available. Similarly, mappings for object properties that are the inverse of some mapped object property are not included in our primary source alignment files. However, we provide an additional, derived file that includes all entailed mappings generated by a reasoner.

### Evaluation and Quality Control

A SPARQL query was developed for automatically verifying the **Totality** of the combined alignments as described in Criterion 4. The query finds any PROV-O class or object property term such that it, or its inverse, is *not* transitively related via equivalence or subsumption, or related via a property chain axiom or SWRL rule, to some BFO, RO, or CCO class or object property term. This query was added to the continuous development pipeline, and generated a list of unmapped terms when triggered. The resulting list of unmapped terms measured our progress during the course of the project.

The SPARQL query specifies each type of mapping relation described in Criterion 1. First, mappings transitively entailed by equivalence or subsumption, as explained above, were excluded using a SPARQL property path expression. Second, entailed mappings from inverse object properties were materialized (storing the result of an intermediate query) into a temporary graph using the HermiT reasoner [22, 23] and ROBOT command-line tool [34] before running the query. Third, mappings involving either an OWL property chain axiom or SWRL rule were also found by the SPARQL query, and excluded from the generated list of unmapped terms.

To test the **Coherence** and **Consistency** of each alignment, as in Criterion 2, every canonical example instance from the W3C documentation for PROV-O and its extensions was copied into RDF files serialized in the Terse Triple Language (TTL or "RDF Turtle") [35] and imported into the editor's module. 312 instances were counted, including either named or anonymous individuals represented with blank nodes. The HermiT reasoner tested the consistency of these example instances with PROV, BFO, RO, CCO, and the alignments between them. HermiT determines whether a set of OWL 2 DL axioms and assertions are satisfiable (i.e. coherent) and consistent by using the hypertableau calculus to explore whether any derivations lead to a contradiction. Two examples were discovered to be inconsistent with PROV-O itself, while two others were inconsistent with our PROV-BFO alignment. These few inconsistencies are mostly due to unintentional, technical errors and described later in the Results section. In our view, this is "a feature, not a bug": The alignments can be used to find inconsistencies in data interpreted as instances of PROV-O classes.

The alignments were further tested for consistency with an alignment between PROV-O and the SOSA (Sensor, Observation, Sample, and Actuator) ontology [36], which is a core subset of the Semantic Sensor Network Ontology (SSN) [37], another W3C recommendation ontology. No inconsistencies were found when testing our alignments with this alignment plus example SOSA instances from its W3C documentation. While our alignments provide additional interoperability between BFO, CCO, RO and SOSA, we reserve a more complete discussion of SOSA for another time.

**Conservativity** of each alignment, as in Criterion 3, was tested by constructing the approximate deductive differences [30] between each set of target ontologies and the union of aligned ontologies. First, all subsumptions and equivalence relations were materialized from each set of target ontologies, and then the aligned ontologies, using HermiT via ROBOT. Second, a SPARQL CONSTRUCT query filtered these materialized relations to those which relate terms from the same ontology (thereby excluding mappings). Third, the ROBOT "diff" command (https://robot.obolibrary.org/diff) compared the output of these SPARQL queries. We found no difference in subsumptions or equivalences between terms within each ontology, which means that the hierarchies of each input ontology were not changed by the alignments. Although this level of conservativity was achieved, our alignment does entail new disjoint relations between PROV-O terms, as we explain in our mapping of PROV Agent.

Automated tests for the techniques described above were implemented as part of an ontology engineering pipeline using ROBOT and GNU Make (https://www.gnu.org/software/make/). ROBOT commands used for running SPARQL queries and the HermiT reasoner were composed into Makefile tasks. These Makefile tasks are run within a continuous development pipeline using GitHub Actions. The result is that changes to the alignments can be automatically, rigorously tested when committed to the Git repository hosted in GitHub.

To narrow down possible object property mappings, a SPARQL query was developed to find BFO, RO, and CCO object properties with the same (possibly inherited) domain and range as each PROV object property. For example, PROV *generated* has domain PROV Activity and range PROV Entity. When these classes were mapped as subclasses of BFO process and BFO continuant, respectively, the SPARQL query found three object property matches: CCO *affects*, CCO *has input*, and CCO *has output* all have BFO process and continuant as their domain and range. The last, CCO *has output*, was then manually selected by curators because it best matches the meaning of PROV *generated*. This semi-automated method of object property matching by narrowing down candidate mappings is especially useful when the domain or range of an object property is not always obvious based on its label.

### FAIR Compliance

The technical artifacts produced by our work comply with FAIR (Findable, Accessible, Interoperable, Reusable) principles [19]. All artifacts, data, and code are maintained in a public GitHub repository (https://github.com/BFO-Mappings/PROV-to-BFO) and can also be found and accessed on Zenodo (https://doi.org/10.5281/zenodo.14692262) [18]. We look forward to adding our alignments to relevant registries such as OBO Foundry [13] or an SSSOM mapping commons [7]. To support interoperability and reuse, the alignments are implemented in OWL 2 DL and serialized in RDF Turtle. They are maintained in versioned, annotated files separate from the corresponding ontologies, similar to the Dublin Core extension of PROV-O [38]. Distributing separate files, as opposed to changing the corresponding ontologies, preserves compatibility with their existing usage.

The alignments can be used importing them into any OWL ontology which uses terms from BFO or PROV-O ontologies. To benefit from the full logical implications of the alignments, PROV-O and BFO should also be imported. Our alignment files do not import these ontologies directly in order to maximize modularity and reuse. An example usage file is included in the Github repository.

Provenance metadata for the alignments are encoded as annotations on each OWL axiom that represents a mapping relation. In RDF Turtle, the mapping relations themselves are serialized with **OWL *annotatedSource***, **OWL *annotatedProperty***, and **OWL *annotatedTarget*** properties of reified OWL axioms [39], which correspond to the subject, predicate, and object of a mapping, respectively. These axioms are interpreted like any other OWL class or object property axiom despite being syntactically encoded in an indirect way. SWRL rule mappings are similarly serialized. Informal justifications for each mapping are annotated using RDFS *comment*. Although automated ontology matching techniques typically include a confidence value or similarity measure for each mapping, we did not include these because our mapping process was based on manually analyzing the meaning of each term. An example of a mapping encoded in RDF Turtle is shown below:

```
[]  rdf:type owl:Axiom ;
    owl:annotatedSource   prov:Activity ;
    owl:annotatedProperty owl:equivalentClass ;
    owl:annotatedTarget   obo:BFO_0000015 ;
    sssom:object_label    "process" ;
    rdfs:comment "A prov:Activity is equivalent to a process because it happens over time, while not being a temporal region itself."@en .
```

Annotation properties from the Simple Standard for Sharing Ontology Mappings (SSSOM) [7] vocabulary were also used. The SSSOM annotation properties *subject label* and *object label* provide convenient, human friendly references for terms with opaque IRIs. For example, the BFO term labeled "process" has the IRI ending `BFO_0000015`, so an axiom which maps that term can be annotated with that label. Although SSSOM as a broader standard recommends maintaining mappings in a tabular format, its tooling can export to annotated, reified OWL axioms (https://mapping-commons.github.io/sssom/spec-formats-owl/). We therefore chose to serialize our mappings directly in OWL and then used SPARQL to generate a SSSOM-compatible CSV file to avoid being dependent on SSSOM. More importantly, the tabular SSSOM schema and tooling currently does not currently support a clear standard for complex mappings. Each of our non-complex mappings in the CSV is annotated with the SSSOM *mapping justification* of "manual mapping curation" from the Semantic Mapping Vocabulary (SEMAPV) [40].

As of this publication, our alignments use PROV-O version 2013-04-13, BFO 2020 version 2024-01-29, CCO version 2024-11-06, and RO version 2024-04-24. Each alignment is annotated with its version and the versions of each ontology it targets using PROV *wasDerivedFrom*. Our PROV-BFO, PROV-CCO, and PROV-RO alignments are encoded in separate files with some redundancy to make them independently usable. For example, PROV Plan is mapped as a subclass of BFO generically dependent continuant in the BFO mapping, while also mapped as a subclass of CCO Information Content Entity in the CCO mapping, where the latter entails the former. This allows for use of the BFO mapping independently of the CCO mapping. RO and CCO alignments include OWL imports of the BFO alignment because RO and CCO logically depend on BFO. Lastly, a separate RDF Turtle file imports all three of the alignment files, along with the BFO, RO, and CCO ontologies, for viewing the alignments in context in Protege [41] and for testing with reasoners and SPARQL queries.

## Results

A total alignment in the sense of Criterion 4 was achieved by mapping all 153 classes and object properties in PROV-O and its extensions using equivalence and subsumption relations defined in Criterion 1. The alignments were verified to be coherent and conservative in the senses defined in Criteria 2 and 3. Consistency of the alignments with all but two canonical PROV-O examples was verified, which we believe is due to minor errors in the PROV-O documentation. We were unable to encode complex mappings for data properties into a computable format, but we explain them informally below. Most mappings are implicit, requiring a semantic reasoner to materialize them. 35 terms were explicitly mapped to BFO using 6 equivalence relations, 24 subsumption relations, and 8 SWRL rules. 37 terms were explicitly mapped to CCO using 5 equivalence relations, 23 subsumption relations, 1 property chain, and 6 SWRL rules. Finally, 25 terms were explicitly mapped to RO using 26 subsumption relations. Among all mappings, 4 terms were supplemented with additional SKOS mappings to provide commentary on possible alternative mappings. Implicit mappings for all other PROV-O terms are entailed by logical axioms of the aligned ontologies. Some class mappings are displayed in **Figure 1** and **Figure 2**.

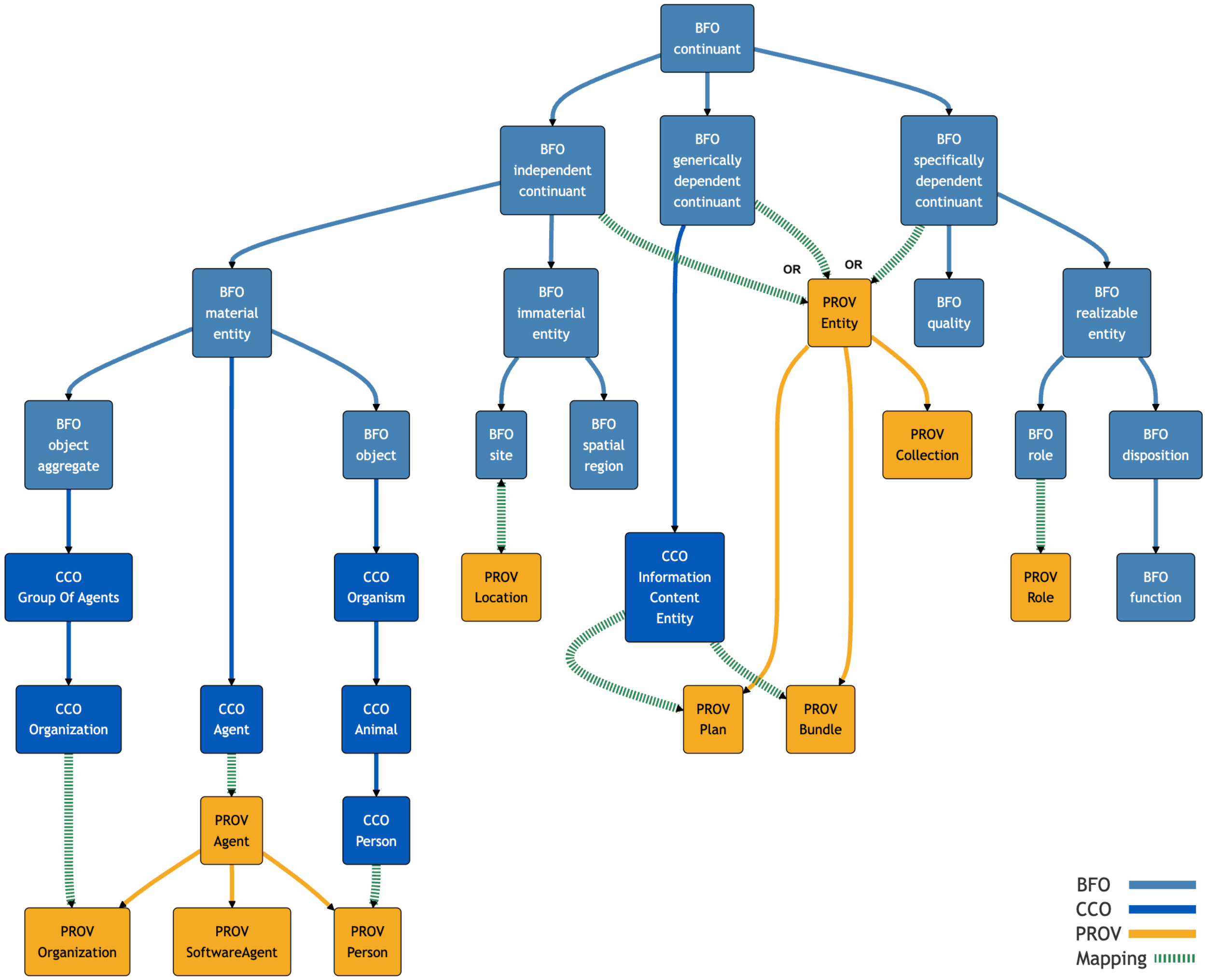

**Figure 1**: Simplified illustration of mappings between PROV-O and BFO continuants, shown with dashed arrows. Bidirectional arrows show equivalence mappings. Solid arrows show the original hierarchies within each ontology. This does not show all complex relationships between terms represented with OWL axioms. Note that some apparently homonymous terms are not semantically equivalent, such as “agent”, “person” and “organization”.

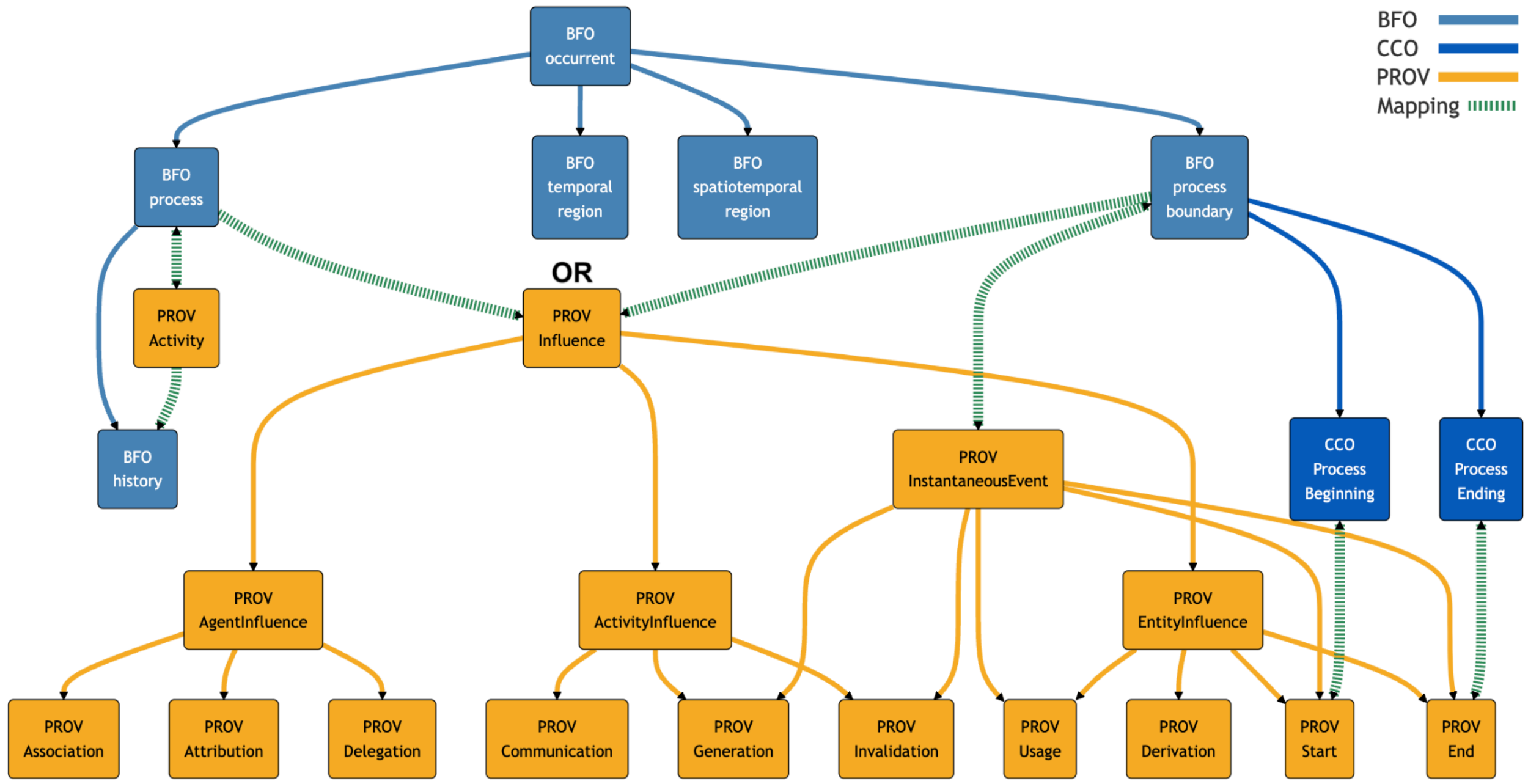

**Figure 2**: Simplified illustration of some mappings between PROV-O and BFO occurrents. This does not show the disjoint relationship between BFO process and BFO process boundary. Instances of PROV Influence can be either processes or process boundaries, but never both. Note that BFO history is entailed to be a subclass of PROV Activity because PROV Activity is mapped as equivalent to BFO process.

Each mapping is described in order of groupings used by the W3C PROV-O documentation. *Starting Point* terms are referred to as "a small set of classes and properties that can be used to create simple, initial provenance descriptions". *Expanded terms* are subclasses and subproperties of Starting Point terms and also terms used to provide more specific details about provenance. *Qualified terms* are those used to provide "additional attributes of the binary relations" asserted by use of the Starting Point and Expanded terms.

### Starting Point classes and object properties

**PROV Entity** is defined as "a physical, digital, conceptual, or other kind of thing with some fixed aspects; entities may be real or imaginary". Our interpretation of this is that a PROV Entity can exist entirely at different times and persist its identity over time with all spatial parts and no temporal parts. We therefore map PROV Entity as a subclass of **BFO continuant** with one exception. BFO spatial regions are continuants that neither participate in processes nor bear qualities. This exception is captured by mapping PROV Entity to a subclass of things that are **independent continuants** and **not spatial regions**, in a union with **generically dependent** and **specifically dependent continuants** in BFO. In other words, every instance of a PROV Entity is one of these kinds of continuants but not a spatial region. This axiom ensures that any other subclass of BFO continuant, which may be added by a domain ontology, is not automatically entailed to be a superclass of PROV Entity. Structurally, this mapping is supported by the disjoint relationship between PROV Entity and PROV Activity. This disjoint relationship is entailed by additionally mapping PROV Activity to a subclass of BFO occurrent because continuants and occurrents are disjoint.

**PROV Agent** is mapped as a subclass of **BFO material entities** that both ***participate in*** some PROV Activity and ***bear*** some **BFO role** that is ***realized in*** a PROV Activity (see **Figure 3**). The reason for this mapping is that a PROV Agent always has some matter as a part that persists in time. This is true even for instances of its subclass, PROV SoftwareAgent, which is defined as "running software", because every particular instance of a PROV SoftwareAgent is a material realization of some software (which may itself be considered a generically dependent continuant). The definition of PROV Agent also states that an agent "bears some form of responsibility for an activity taking place, for the existence of an entity, or another agent's activity". This is formalized in axioms which entail that every PROV Agent *participates in*, and *bears* some role that is *realized in*, some PROV Activity. According to the "open-world" [42] interpretation of OWL 2, this does not necessarily imply that any data set about a PROV Agent should also include data about some PROV Activity or BFO role, but rather only asserts the existence of those related entities. According to Requirement VI3 in "The Rationale of Prov" [43], a PROV Agent could be a PROV Entity and this is possible in our mapping because both PROV Agent and PROV Entity are subclasses of BFO continuant.

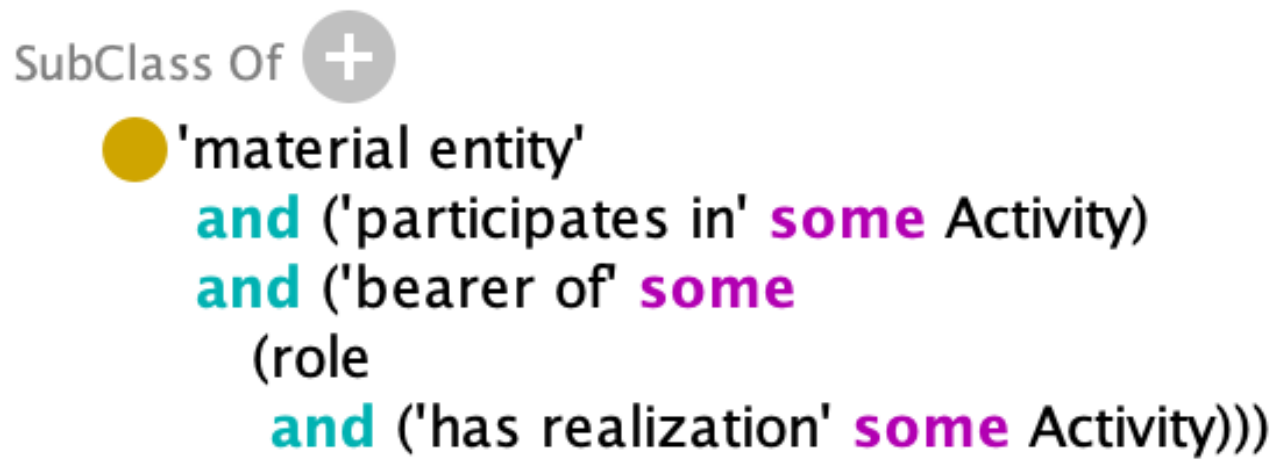

**Figure 3:** PROV Agent is a subclass of BFO material entities that are related to PROV activities in certain ways (visualized in Protege).

Having created a subsumption to a general class in BFO, we then developed a more specific, equivalence mapping for PROV Agent in CCO: A **PROV Agent** is equivalent to the intersection of **CCO agents** that are a CCO ***agent in*** some PROV Activity. The BFO and CCO mappings are both provided for BFO users who are not CCO users, and for CCO users who want to benefit from the more specific equivalence relation mapping.

**PROV Activity** is mapped as equivalent to the class **BFO process**. The definition of PROV Activity includes "something that occurs over a period of time and acts upon or with entities". Similarly, a BFO occurrent - the parent class of BFO process - is defined as "an entity that unfolds itself in time or it is the start or end of such an entity or it is a temporal or spatiotemporal region". However, instances of PROV Activity do not seem to include instances of BFO temporal regions, such as the year 1986. Hence, BFO process was selected as a more appropriate equivalence class. Note that, although all instances of PROV Agent are *participants in* some PROV Activity at some time, *not all* instances of PROV Activity have participants. The definition of BFO process says that every process "has some material entity as participant" [12] but this is not formalized as a logical axiom, so we did not include this axiom for PROV Activity either.

An important caveat to our mapping of PROV Agent and PROV Activity relates to *Requirement VI4* in "The Rationale of PROV", which states: "prov is to allow agents to be activities".

This stipulation is described as a consequence of *Requirement GE1*: "prov is to minimize class disjointness constraints and to use strong rationale when defining such constraints" and *Requirement GE2*: "prov is to include the mirror of each concept, where relevant". The authors elaborate: "As a result, being an agent is not an intrinsic characteristic of an entity or activity. Instead, it is the very presence of responsibility relations that implies that some entities or activities are also agents".

It is possible to formally accommodate Requirement V14, by mapping PROV Agent as the disjoint union of (non-spatial region) continuants and processes. This would ensure that every PROV Agent is an instance of a continuant or process but not both. Despite this mapping being a possible implementation, we decided not to include it, as this would ultimately go against the spirit of BFO's core axiom that continuants and occurrents are disjoint. Our mapping is not merely a translation but is also intended to enrich the semantics of each ontology. Mapping PROV Agent to BFO continuant and PROV Activity to BFO process implies that PROV Agent is disjoint with PROV Activity. While this contradicts Requirement V14, it does not contradict any specific example provided in the essay or W3C documentation of something that is both a PROV Agent and PROV Activity.

In this case, there is indeed a strong reason for making continuants and occurrents disjoint. This rationale is given for *Requirement VI7*: "prov is not to allow an activity to be an entity", where the authors state: "An activity represents something that "happened", whereas an entity is a thing, whether real or imaginary. This distinction is similar to that between "continuant" and "occurrent" in logic" [43]. Moreover, a simpler explanation of how an activity can be related to some responsibility is that an activity may involve an agent that bears that responsibility. This is formalized by our logical axiom stating that every PROV Agent bears some BFO role that is realized in a PROV Activity. However, this does not and should not entail that a PROV Agent could be a PROV Activity.

We now highlight some Starting Point object properties in PROV, visualized in **Figure 4**. Since the domain and range of **PROV *wasGeneratedBy*** are PROV Entity and PROV Activity, respectively, it is mapped as a subproperty of **BFO *participates in***, whose domain and range are non-spatial region continuants and process, respectively. A more specific subproperty mapping to CCO is also provided: **CCO *is output of*** has domain and range of continuant and process, respectively. Despite redundancy with BFO, a mapping to RO is also provided: **RO *participates in*** has domain and range of continuant and occurrent, respectively. All three mappings are logically consistent when used together. The difference between the RO mapping and BFO mapping is that the latter excludes spatial regions from being used in a PROV *wasGeneratedBy* relation. Both *participates in* predicates were used for practical reasons. While the RO predicate is more compatible with some OBO Foundry ontologies, the BFO predicate is a more refined version.

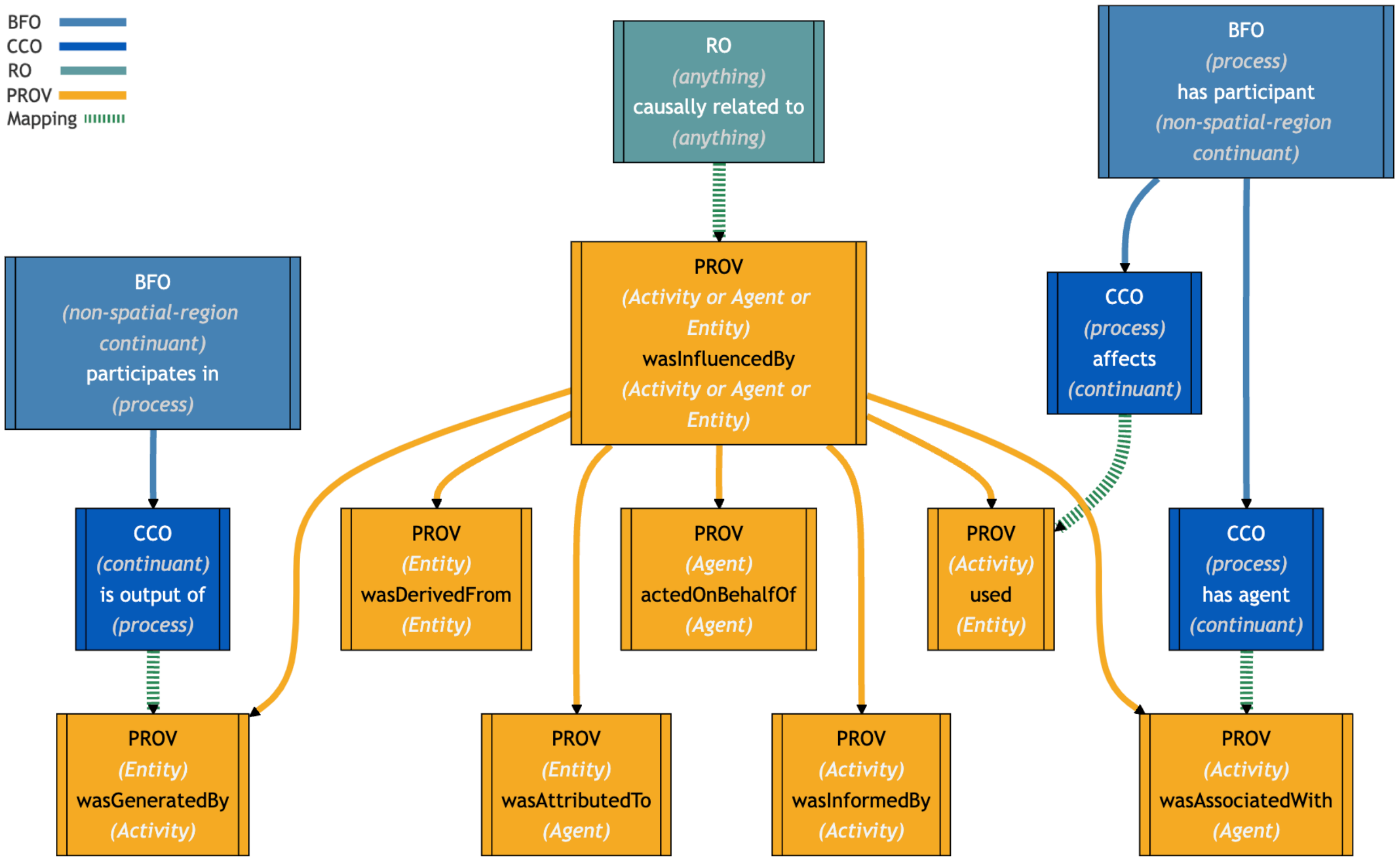

**Figure 4:** Simplified illustration of some PROV-O object property mappings. Solid arrows show relations within an ontology. The domain and range of each property are shown in parentheses. Most object properties and their inverses are not shown here.

**PROV *wasAssociatedWith*** has domain PROV Activity and range PROV Agent. This is used to represent, for example, that an activity of illustrating was associated with an illustrator. This object property is mapped as a subproperty of **BFO *has participant***, which is the inverse of BFO *participates in*. It is also mapped as a subproperty of **CCO *has agent***, which is informally defined as relating between a BFO process and CCO Agent, but does not formally specify these classes as the domain and range. Finally, PROV *wasAssociatedWith* is also mapped to **RO *has participant***, which is the inverse of RO *participates in*.

## Expanded classes and object properties

Among the Expanded classes for PROV-O are PROV Person and PROV Organization. These are good examples of terms which might be incorrectly mapped if relying on automated lexical matching. CCO Person has the same RDFS *label* as PROV Person. However, PROV Person is a subclass of PROV Agent, while CCO Person is not a subclass of CCO Agent. We map **PROV Person** as equivalent to the **intersection** of **CCO Person** and **PROV Agent**. This entails that every PROV Person is both a CCO Person and PROV Agent. Conversely, anyone who is both a CCO Person and a PROV Agent is therefore a PROV Person. However, this does *not* entail that every CCO Person is a PROV Person, much less a PROV Agent. All of the same considerations apply to PROV Organization, which is a subclass of PROV Agent, while CCO Organization is not a subclass of CCO Agent. Therefore, **PROV Organization** is equivalent to the **intersection** of **CCO Organization** and **PROV Agent**. Some of these relationships are illustrated in **Figure 5**.

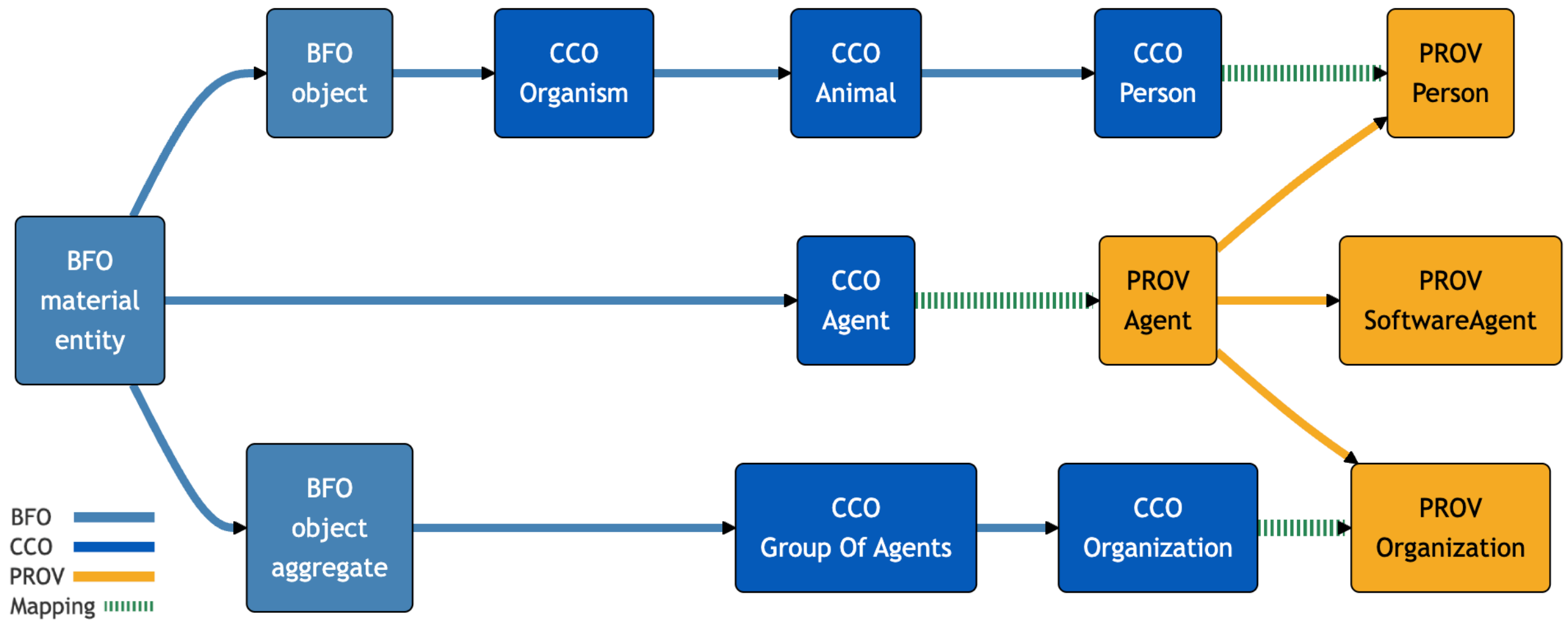

**Figure 5:** Simplified illustration of agents, persons, and organizations in PROV-O and CCO. Note that while CCO Group of Agents is a subclass of BFO object aggregate, CCO Agent is *not* a subclass of BFO object. CCO allows for the possibility that an organization, as a group of agents, could also be a singular agent (e.g. a sports team). However, this does not commit CCO to the possibility that something can be both an object and object aggregate.

**PROV Bundle** is defined as "a named set of provenance descriptions" [8] that "constitute islands of provenance information" [9]. It is intended for annotating metadata about RDF documents, such as the time when a document was generated. We therefore map it as a subclass of **CCO Information Content Entity** and also **BFO generically dependent continuant**. By contrast, we did not provide a more specific mapping for **PROV Collection** because it is defined more generically as "an entity that provides a structure to some constituents, which are themselves entities" [8]. Although some subclasses of PROV Collection like PROV Dictionary are described as data structures [44], we did not want to restrict the kind of continuants that a PROV Collection may be composed of.

**PROV Location** is mapped as equivalent to **BFO site**, which is defined as "a three-dimensional immaterial entity whose boundaries either (partially or wholly) coincide with the boundaries of one or more material entities or have locations determined in relation to some material entity". On the other hand, the informal definition of PROV Location states that an instance of it "can also be a non-geographic place such as a directory, row, or column". This fits with our mapping if we interpret a row in a digital spreadsheet, for example, as being located inside and bound by some material realization of a computer. However, an alternative, broader mapping of PROV Location to another subclass of BFO continuant is also possible.

Expanded object properties include **PROV *hadPrimarySource***, ***wasQuotedFrom***, and ***wasRevisionOf***. These are all subproperties of **PROV *wasDerivedFrom***, with domain and range of PROV Entity, which is mapped as a subproperty of **RO *causally influenced by***. This mapping allows us to interpret any derivation relation represented in PROV-O as a causal relation between entities. By contrast, **PROV *wasInvalidatedBy*** has domain and range of PROV Entity and PROV Activity, respectively. This was mapped as a subproperty of both **BFO *participates in*** and **RO *participates in***, which have slightly different domains and ranges, as mentioned previously. Finally, PROV *wasInvalidatedBy* is also mapped as a subproperty of **CCO *is affected by***, which has BFO continuant and process as its domain and range.

**PROV *atLocation*** is mapped using SWRL rules. Its domain is the union of PROV Agent, PROV Activity, PROV Entity and PROV InstantaneousEvent, while its range is PROV Location. In BFO, there are two object properties similar to PROV *atLocation*. The domain of **BFO *occurs in*** is the union of BFO process and process boundary, while its range is the union of BFO site and material entity. **BFO *located in*** has a domain and range of BFO independent continuants that are not spatial regions. In order to map PROV *atLocation* to both of these object properties, SWRL rules were used to restrict the domain and range of particular instances of them. Here are the SWRL rule mappings between PROV *atLocation* and BFO *occurs in* (`obo:BFO_0000066`), preceded with informal descriptions of each:

*"If a PROV Activity is **at** some **location**, then it **occurs in** that location"*

```
prov:atLocation(?x,?y) ^ prov:Activity(?x) ->
obo:BFO_0000066(?x,?y)
```

*"If a PROV InstantaneousEvent is **at** some **location**, then it **occurs in** that location"*

```
prov:atLocation(?x,?y) ^ prov:InstantaneousEvent(?x) ->
obo:BFO_0000066(?x,?y)
```

*"If something **occurs in** a PROV Location, then it is **at** that **Location**"*

```
obo:BFO_0000066(?x,?y) ^ prov:Location(?y) ->
prov:atLocation(?x,?y)
```

These rules restrict the domain of PROV *atLocation* to the domain of BFO *occurs in*, the union of process and process boundary, which are mapped as equivalent to PROV Activity and PROV InstantaneousEvent, respectively. This allows a semantic reasoner to automatically infer that some subset of the instances of the PROV *atLocation* relation are also instances of the BFO *occurs in* relation – only those involving a PROV Activity or PROV InstantaneousEvent. Conversely, the last rule restricts the range of *occurs in* to entail that anything related by the *occurs in* relation to a PROV Location is also related to it by the PROV *atLocation* relation. These rules are necessary because some instances of PROV *atLocation* could involve a PROV Entity or PROV Agent, which are disjoint with BFO process and process boundary in virtue of being subclasses of BFO continuant. That is, instances of PROV *atLocation* can be inferred to be instances of BFO *occurs in* only when involving instances of PROV Activity or PROV InstantaneousEvent. Similar SWRL rules provide a mapping between PROV *atLocation* and BFO *located in*.

### Qualified classes

The Qualification Pattern employed in PROV-O is a form of *reification*, which is generally recommended as a standard methodology for alleviating the representational limits of description logic [45] and representing *n*-ary information among the OWL community [46]. To associate further information to a binary unqualified relation of "influence" between entities, activities or agents, the Qualification Pattern reifies this relation as an instance of the class PROV Influence. The entities, activities, and agents involved become related to the PROV Influence via subproperties of PROV *qualifiedInfluence* or PROV *influencer*, depending on whether they are the subject or object of the influence relation, respectively. Additional properties can then be added to the PROV Influence, such as when it occurred. For example, the RDF triple `:sortActivity prov:used :datasetA` could be qualified as an instance of `prov:Usage` (see **Figure 6**) and then annotated with `prov:atTime`. Note that while we mapped terms involved with the Qualification Pattern, our alignments are not intended to force users to adopt this pattern or reification more generally. For representing *n*-ary information, some alternative options are the description/situation strategy [47], the strategy of non-rigid classification [48], and the strategy of temporally qualified continuants [49].

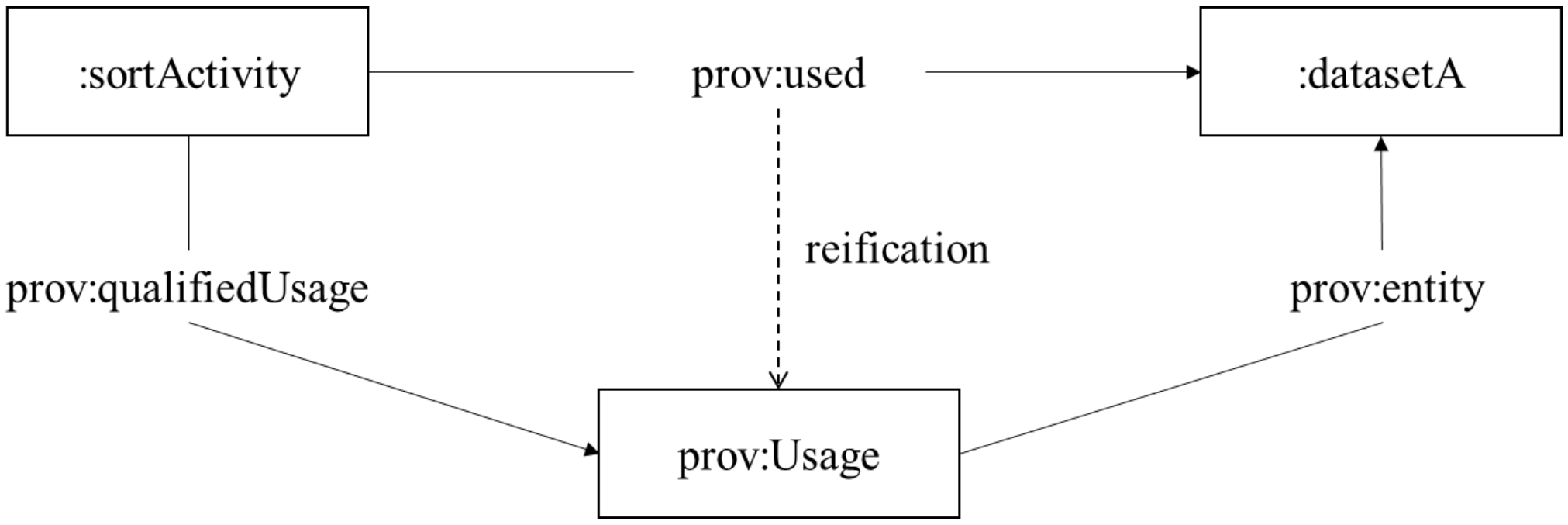

**Figure 6:** Illustration of a Qualified Pattern [8].

**PROV Influence**, as the superclass of 16 Qualified Influence classes, is mapped to a subclass of the *disjoint union* of **BFO process** and **BFO process boundary**. Some of its subclasses such as **PROV Generation**, **PROV Start**, and **PROV End** are subsumed under **PROV InstantaneousEvent**, which is equivalently mapped to **BFO process boundary** since instances of PROV InstantaneousEvent are indivisible boundaries of some PROV Activity that is equivalent to a BFO process. Other subclasses of PROV Influence, like **PROV Communication** and **PROV Derivation**, are not subsumed under PROV InstantaneousEvent, and therefore are mapped as subclasses of **BFO process** instead. Either way, however, no instance of a PROV Influence subclass can be both a process and a process boundary because these classes are disjoint in BFO. See **Figure 2** for simplified illustrations of these mappings.

It is worth noting a radical divergence of our mapping of PROV Influence from its original definition in PROV-O. According to its original definition,

> "Influence is the capacity of an entity, activity, or agent to have an effect on the character, development, or behavior of another by means of usage, start, end, generation, invalidation, communication, derivation, attribution, association, or delegation."

In our view, PROV-O authors' conception of influence as capacity is intended to emphasize the dependent nature of PROV Influence. After all, a capacity, understood as a special kind of BFO disposition, would be a BFO specifically dependent continuant since it specifically depends on some independent continuant as its bearer. However, the dependent nature of PROV influences is better explained by classifying these as BFO processes or process boundaries that depend on agents through relations such as BFO *has participant*. Given our alternative conception, a PROV Influence may *realize* a capacity as BFO disposition, but it is not thereby identical to that capacity, nor is anything like this represented in any canonical PROV-O example.

Take an example from the W3C documentation, where the illustration and derivation of a bar chart are both classified as examples of PROV Influence ([https://www.w3.org/TR/prov-o/#hadGeneration](https://www.w3.org/TR/prov-o/#hadGeneration)) (see **Figure 7**). Here an act of illustrating a bar chart, when considered as a type of influence on the bar chart, clearly depends on the existence of an illustrator. The bar chart could also be derived from some data set, and this derivation clearly would depend on the data set. However, in both cases, it seems more appropriate to consider these influences as *events* occurring in time and not *capacities* of the illustrator or data set, respectively.

```
:bar_chart
   a prov:Entity, ex:Chart;
   prov:wasDerivedFrom   :aggregatedByRegions;
   prov:qualifiedDerivation [
      a prov:Derivation;
      prov:entity        :aggregatedByRegions;
      prov:hadGeneration :illustration;
   ];
.

:aggregatedByRegions a ex:Dataset .

:illustration
   a prov:Generation,
     prov:InstantaneousEvent;
   prov:activity :illustrationActivity;
   prov:atTime   "2012-04-03T00:00:11Z"^^xsd:dateTime;
.

:illustrationActivity
   a prov:Activity;
   prov:startedAtTime "2012-04-03T00:00:00Z"^^xsd:dateTime;
   prov:endedAtTime   "2012-04-03T00:00:25Z"^^xsd:dateTime;
.
```

**Figure 7:** Example instances of PROV Influence represented in RDF Turtle from the W3C documentation [8]. These do not represent "capacities" of an agent but instead a specific events in time – the derivation of a bar chart and the generation of that bar chart during an illustrating activity – which an agent may have nevertheless been causally responsible for.

Moreover, this example provides a further reason why not all influences are capacities. Note that some PROV influences, like the generation of a bar chart, only exist at an instant of time. Assuming that no capacity only exists at an instant of time, then our mapping of PROV InstantaneousEvent to BFO process boundary entails that some instances of PROV influences, such as PROV generations, cannot be classified as capacities as BFO dispositions. This is because process boundaries and dispositions are disjoint in BFO: the former are occurrents while the latter are continuants.

PROV-O further divides PROV Influence into 3 direct subclasses – **PROV EntityInfluence**, **PROV ActivityInfluence**, **PROV AgentInfluence** – according to a trisection of influencers. Since an influencer is an entity, activity, or agent, the influence it exerts is, accordingly, an entity influence, activity influence, or agent influence. These 3 influence classes lack equivalent mappings in BFO, but they are automatically mapped into a subclass of the disjoint union of BFO process and BFO process boundary since they are subclasses of PROV Influence.

CCO is used to provide two additional cases of equivalent mappings. **PROV Start** is equivalently mapped to **CCO process beginning**, while **PROV End** is equivalently mapped to **CCO process ending**. PROV-O treats PROV Start and End as instantaneous events connected to particular activities via the Qualification Properties PROV *qualifiedStart* and *qualifiedEnd*, subject to further restrictions by particular triggering entities, locations, time data, and so on. So, PROV Start and PROV End can be mapped to CCO process beginning and CCO process ending. A process beginning/ending is a process boundary occurring on the starting/ending instant of the temporal period, which is occupied by a process that is equivalent to a PROV Activity.

**PROV Role** is defined as "the function of an entity or agent with respect to an activity, in the context of a usage, generation, invalidation, association, start, and end". Although the word "function" is used here, BFO distinguishes between roles and functions. While a BFO function must be partially internally determined by the physical make-up of its bearer, a BFO role could be completely externally determined and can be gained or lost by its bearer. For example, a hammer's function to drive nails is based on its shape, but a hammer can also play the role of a paper weight when necessary. In light of the BFO function/role distinction, we map PROV Role directly as a subclass of **BFO role** on the grounds that a PROV Role is externally determined by the context in which the bearer plays the role. However, PROV Role is not equivalent to BFO role because it is defined as existing specifically in a "context of a usage, generation, invalidation, association, start, and end".

The case of **PROV Plan** is more complicated. It is mapped to a subclass of **CCO Information Content Entity**. According to CCO, an Information Content Entity is a BFO generically dependent continuant (GDC) that generically depends on some CCO Information Bearing Entity and stands in relation of aboutness to some entity [17]. It is clear that a PROV Plan is a GDC because it may have multiple copies and thus does not have to depend on any specific bearer. And a PROV Plan is clearly *about* some entity in the BFO or CCO sense since it "represents a set of actions or steps intended by one or more agents to achieve some goals" [8]. Also, given that CCO Information Content Entity is a subclass of BFO generically dependent continuant, our CCO mapping result implies that PROV Plan is mapped to a subclass of GDC in BFO.

However, it is worth noting that PROV Plan is not mapped to **CCO Plan**, though they are similar to some degree. A CCO Plan is a Directive Information Content Entity that prescribes some set of intended CCO Intentional Acts through which some Agent expects to achieve some CCO Objective. Here CCO Plan is constrained to the prescription of intentional acts, but PROV Plan is allowed to prescribe activities or processes that are broader than intentional acts. As its definition reveals, "[t]here exist no prescriptive requirement on the nature of plans, their representation, the actions or steps they consist of, or their intended goals" [8]. Therefore, there is no reason for limiting PROV Plan only to Directive Information Content Entity, and thus there is no equivalent mapping between PROV Plan and CCO Plan.

### Qualified object properties

PROV-O contains 14 Qualification Properties, 4 Influencer Properties, and 6 additional object properties. **PROV *qualifiedInfluence*** and **PROV *influencer*** are the superproperties of Qualification properties and Influencer properties, respectively. Given their maximal causal generality, a plausible proposal is to appeal to RO by mapping them both as subproperties of **RO *causally related to***. The same is for **PROV *wasInfluencedBy***, the most general one among the 6 additional qualified properties.

We highlight mappings of PROV-O qualified properties whose domain or range is mapped to a subclass of BFO process boundary. Few object properties in BFO, RO, and CCO directly involve process boundary. This fact also indirectly shows why there exists no equivalent mapping for some qualified properties. For example, **PROV *qualifiedStart***, ***qualifiedEnd***, and ***qualifiedUsage*** take PROV Activity as their shared domain, and their respective ranges are PROV Start, PROV End, and PROV Usage, which are all mapped as subclasses of BFO process boundary. We thus find it tractable to map their domains and ranges to the parent class of BFO process boundary: occurrent. In light of this, the most appropriate way is to map them as subproperties of **BFO *has temporal part***, because it takes occurrents including process boundaries as their domain and range. Informally, the start and end of an activity, and an activity's usage of an entity, are simply temporal parts of an activity.

A more difficult type of cases concerns how to map **PROV *qualifiedGeneration*** and **PROV *qualifiedInvalidation***. As their respective range, PROV Generation and PROV Invalidation are also mapped as subclasses of BFO process boundary. Given that their shared domain is PROV Entity, however, they cannot be mapped as subproperties of **BFO *has temporal part***. On the other hand, the meanings of PROV *qualifiedGeneration* and PROV *qualifiedInvalidation* are somewhat similar to those of **CCO *is output of*** and **CCO *is affected by***, respectively. But it is still inconsistent to directly map the former to the latter by the subsumption relation because the latter take process as their shared range but process is disjoint with process boundary (as the former's range). Therefore, we have to appeal to SKOS *relatedMatch* as a last resort here. Nevertheless, all of these qualified properties are still transitively subsumed by our mapping of PROV *qualifiedInfluence* to RO *causally related to*.

Up to now we have reviewed all important mapping cases of qualified classes and object properties. The remaining qualified properties, though not discussed here, are encoded in the alignment files [18].

### Data properties

While we considered complex mappings for PROV-O data properties, we were unable to encode these in a computable format due to representational limits of OWL, RDF, and SWRL. For example, the data properties **PROV *startedAtTime*** and ***endedAtTime*** relate a PROV Activity to a `xsd:dateTime` value. In BFO, the relationship between these entities is encoded in a more fine-grained way to make clear the distinctions between a process, the time when it occurred, and the different names for that time. **Figure 8** shows how a PROV Activity, as a process, *occupies* a temporal region with a *first instant*. That instant can be *designated* with an identifier, a **CCO Time of Day Identifier**, which is a BFO generically dependent continuant. In turn, that identifier may be encoded in different ways, such as with a **CCO Document Field**, a BFO independent continuant, that is related to the `xsd:dateTime` value with **CCO *has date time value***. PROV-O's remaining data properties could be similarly translated with even more complex relationships represented in BFO and CCO. It may be possible to represent these complex mappings with SPARQL CONSTRUCT queries.

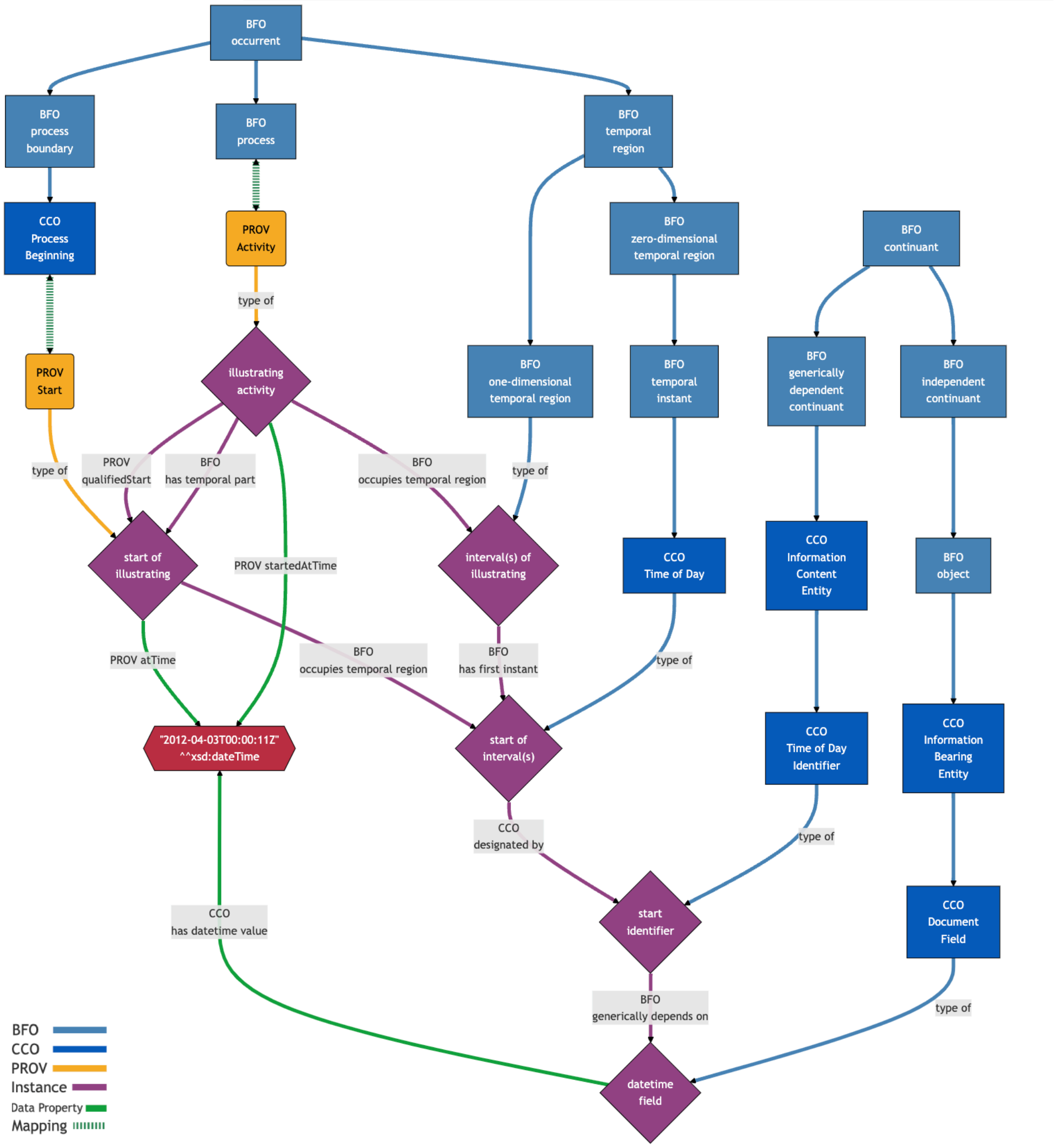

**Figure 8:** Conceptual mapping of PROV *startedAtTime* and *atTime* to several BFO and CCO properties. Data properties are shown in green. Note that PROV-O can similarly encode more information using its reified PROV Start class.

### Inconsistencies

All alignments were tested by importing them, along with BFO, RO, CCO, and all example instances (312 individuals) from the W3C documentation into a temporary OWL ontology that is verified for satisfiability and consistency using the HermiT reasoner. Two examples from the W3C PROV-O documentation were discovered to be inconsistent with PROV-O itself, independently of our alignments. First, *Example 4* (https://www.w3.org/TR/prov-o/#narrative-example-expanded-3) relates an instance of publication activity to a post editor via the PROV *wasAttributedTo* relation, whose domain is PROV Entity. This entails that the publication activity is both a PROV Activity and a PROV Entity, but these classes are disjoint in PROV-O, so this results in a contradiction. Second, the example for PROV Revision relates an instance of PROV Entity to someone via the PROV *wasAssociatedWith* relation, whose domain is PROV Activity. This results in a contradiction because PROV Entity and PROV Activity are disjoint.

Two example instances were found to be inconsistent with our PROV-BFO alignment. One is an instance of PROV Entity, a digested protein sample that is related to a different, (non-digested) protein sample via the PROV *entity* object property (https://www.w3.org/TR/prov-o/#hadUsage) (see **Figure 9**). This object property is used to qualify the influence of one entity on another, and so its domain is PROV EntityInfluence.

```
:digestedProteinSample1
   a prov:Entity;
   prov:wasDerivedFrom :proteinSample;
   prov:qualifiedDerivation [
      a prov:Derivation;
      prov:hadUsage [
         a prov:Usage;
         prov:entity :Trypsin;
        prov:hadRole :treatmentEnzyme;
      ];
   ];
   prov:entity :proteinSample;
.
:proteinSample a prov:Entity .
```

**Figure 9**: A PROV-O example instance that contradicts our PROV-BFO alignment [8]. The digested protein sample, a BFO continuant, is inferred to also be an instance of PROV EntityInfluence, a BFO occurrent.

The above example assertions entail that the digested protein sample is also an instance of PROV EntityInfluence. However, PROV Entity is a continuant while PROV EntityInfluence is an occurrent, and these classes are disjoint. Therefore, the disjointness of continuants and occurrents is inconsistent with the digested protein sample being both a PROV Entity and a PROV EntityInfluence. To explain this inconsistency with our alignment, note that the subject of the PROV *entity* relation should have been the PROV Derivation, not the digested protein sample, as shown in **Figure 10** and similarly in a previous example shown in **Figure 7**. This demonstrates how our alignments can be used to find mistakes that are otherwise consistent with PROV-O.

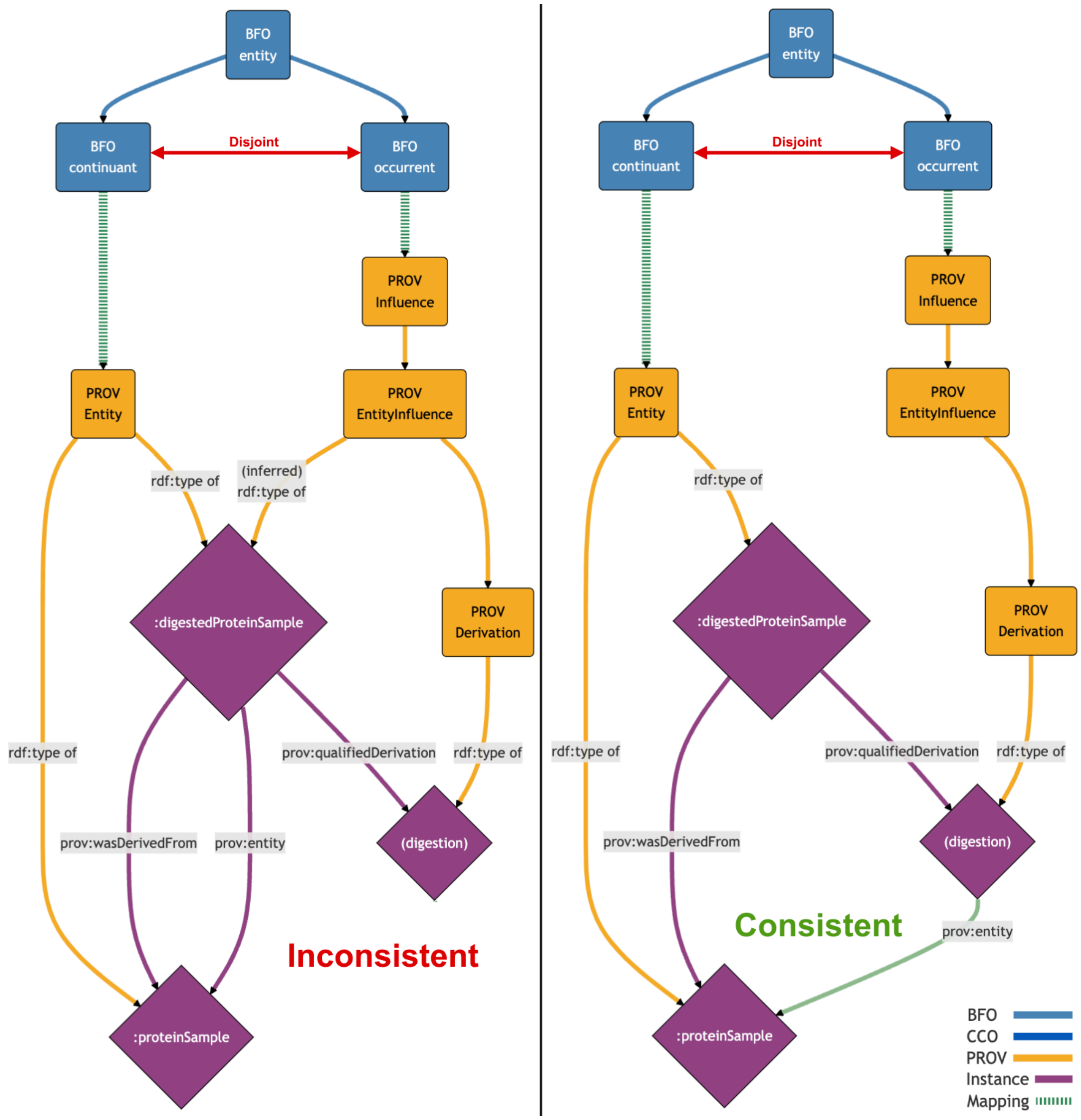

**Figure 10**: Simplified visual model of the digested protein example from Figure 7. Instances are shown as diamonds with purple lines indicating relations between them. The anonymous PROV Derivation instance is named "(digestion)". The left side shows the original example, while the right side shows our suggested correction with the modified relation in green.

Another case of inconsistency concerns an instance of PROV Activity, `:sortActivity`, that is tagged with PROV *atTime* (https://www.w3.org/TR/prov-o/#used) (see **Figure 11**). PROV *atTime* is defined as "the time at which an InstantaneousEvent occurred, in the form of `xsd:dateTime`", having domain PROV InstantaneousEvent, which is equivalently mapped to BFO process boundary. But in our alignment, PROV Activity is mapped to BFO process, which is disjoint with process boundary. So this case is inconsistent with our alignment again. Again, to explain this inconsistency with our mapping, note that this same instance is related in other examples via the PROV *startedAtTime* data property, whose domain is PROV Activity (https://www.w3.org/TR/prov-o/#Usage) (see **Figure 12**). This suggests that this one example may have mistakenly used PROV *atTime* instead of PROV *startedAtTime*.

```
:sortActivity
   a prov:Activity;
   prov:atTime     "2011-07-16T01:52:02Z"^^xsd:dateTime;
   prov:used       :datasetA;
   prov:generated  :datasetB;
.
```

**Figure 11**: An instance of PROV-O that is inconsistent with our PROV-BFO alignment [8]. We believe PROV *startedAtTime*, whose domain is PROV Activity, should have been used here instead of PROV *atTime*, whose domain is PROV InstantaneousEvent.

```
:sortActivity
   a prov:Activity;
   prov:startedAtTime  "2011-07-16T01:52:02Z"^^xsd:dateTime;
   prov:qualifiedUsage [
      a prov:Usage;
      prov:entity    :datasetA;         ## The entity used by the prov:Usage
      prov:hadRole   :inputToBeSorted;  ## the role of the entity in this prov:Usage
   ];
   prov:generated :datasetB;
.
```

**Figure 12**: Another example of the same PROV-O instance that uses a different object property, PROV *startedAtTime*, consistent with our PROV-BFO alignment [8].

## Discussion

A total alignment was achieved such that every class and object property from PROV-O and its extensions is mapped to some class or object property from BFO or its extensions, RO and CCO. Complex data property mappings may be possible in future versions. The mapping predicates used represent simple or complex equivalence and subsumption relations, implemented with OWL *equivalentClass*, OWL *equivalentProperty*, RDFS *subClassOf*, RDFS *subPropertyOf*, OWL *propertyChainAxiom*, and SWRL rules. OWL axiom annotations encode provenance metadata for each mapping. Coherence, consistency, and conservativity of our alignments was verified with a variety of ontology engineering techniques. The alignments serve as a test for these techniques, which may be reused in future projects. SPARQL queries, SWRL rules, OWL axiom annotations, and SKOS metadata are all W3C-affiliated options for ontology mapping in support of FAIR principles.

The current alignments are practically useful for a number of reasons. They allow for integration of PROV-O data in projects using BFO, RO, or CCO, and vice versa. For example, our PROV-BFO alignment entails a SOSA-BFO alignment when combined with PROV-SOSA alignment [36]. Most significantly, the alignments demonstrate how BFO and related ontologies can be used to find inconsistencies in data by enriching it with additional logical axioms. In this case, potential mistakes in example data were found that were consistent with PROV-O alone. The initial set of BFO class mappings was used to narrow down possible object property mappings, which demonstrates a benefit of aligning to upper-level ontologies. Finally, the alignments may serve as a reference alignment for evaluating and improving automated ontology matching systems [6].

Our methodology described stages of alignment to maximize semantic interoperability in terms of the types and number of mapping relations between ontologies. An alignment which provides a full interpretation of PROV-O into BFO-based ontologies using only equivalence relations may be achievable in future work. However, we do not recommend adding terms to either sets of ontologies for the sole purpose of creating equivalence mappings between individual terms. Instead, existing terms should be reused. For example, we did not find an equivalence mapping for PROV *wasAttributedTo*. Users of BFO-based ontologies should use this term with our current alignments instead of creating an equivalent term. An exception to this would be to add terms that are not individually equivalent to an existing term, but may be used for other independent reasons. Reusing existing terms helps ensure that ontologies do not overlap in scope. When an existing term is reused, a mapping is unnecessary and therefore semantic interoperability is already achieved.

## Data availability

The alignment files [18] for each ontology and all related project resources are archived at https://doi.org/10.5281/zenodo.14692262 for use according to the Creative Commons Zero v1.0 Universal license.

## Code availability

The mapping project is maintained in a Git repository hosted in GitHub at https://github.com/BFO-Mappings/PROV-to-BFO. This repository contains the RDF Turtle files, SPARQL query files, GNU Makefile, and Protege catalog file used for representing and verifying the alignments. The project dependencies and their versions are listed in the repository, which include the tools ROBOT, HermiT, GNU Make, and specific versions of RDF, RDFS, OWL, SPARQL, and SWRL.

## Acknowledgements

The authors would like to thank Timothy Lebo for guidance and context about PROV-O. Jonathan Vajda and Ali Hasanzadeh also provided feedback about BFO.

## Author information

### Authors and Affiliations

**National Center for Ontological Research, University at Buffalo, Buffalo, NY, USA**

Tim Prudhomme, Giacomo De Colle, Austin Liebers, Alec Sculley, Peihong "Karl" Xie, Sydney Cohen, & John Beverley

**Department of Philosophy, University at Buffalo, Buffalo, NY, USA**

Giacomo De Colle, Austin Liebers, & John Beverley

**Conceptual Systems, Chapel Hill, NC, USA**

Tim Prudhomme

**Summit Knowledge Solutions, Arlington, VA, USA**

Alec Sculley

**Department of Philosophy, University of Vienna, Vienna, Austria**

Peihong "Karl" Xie

**Institute for Artificial Intelligence and Data Science, Buffalo, NY, USA**

John Beverley

### Contributions

G.D.C., A.L., A.S., P.X., T.P. and S.C. engineered the mappings by analyzing the ontologies in the context of their associated literature and documentation. T.P. formulated the methodology in terms of theoretical alignment criteria and practical engineering techniques for verifying the alignment. T.P. developed and documented the technical artifacts used to encode and test the alignments, in addition to the automated ontology engineering pipeline.

T.P. was the primary author and editor of the manuscript and illustrator of figures while P.X., G.D.C., A.L., A.S. wrote and reviewed significant sections. G.D.C., A.L., and A.S. presented the project for feedback to separate groups within the ontology community. T.P. and S.C. managed the project by scheduling meetings, assigning tasks, and documenting progress. J.B. served as the primary supervisor and senior researcher responsible for conceiving this project.

## Corresponding Author

Correspondence to Tim Prudhomme.

## Ethics declarations

### Competing Interests

None